\documentclass[pra, showpacs, twocolumn, floatfix]{revtex4}
\usepackage{graphicx}
\usepackage{epstopdf}
\usepackage{amsmath, amsfonts, amssymb, bm}
\begin{document}
\title{Phonon statistics in an acoustical resonator coupled to a pumped two-level emitter}
\author{Victor \surname{Ceban}}
\email{victor.ceban@phys.asm.md}
\author{Mihai A. \surname{Macovei}}
\email{macovei@phys.asm.md}
\affiliation{Institute of Applied Physics, Academy of Sciences of Moldova, 
Academiei str. 5, MD-2028 Chi\c{s}in\u{a}u, Moldova}
\date{\today}
\begin{abstract}
The concept of an acoustical analog of the optical laser has been developed recently both in theoretical as well as experimental works. 
We discuss here a model of a coherent phonon generator with a direct signature of quantum properties of the sound vibrations. The 
considered setup is made of a laser driven quantum dot (QD) embedded in an acoustical nanocavity. The system's dynamics is solved 
for a single phonon mode in the steady-state and in the strong QD-phonon coupling regime beyond the secular approximation. We 
demonstrate that the phonon statistics exhibits quantum features, i.e. sub-Poissonian phonon statistics.
\end{abstract}
\maketitle
\section{Introduction}
Sub-Poissonian statistics of vibrational states is a pure quantum propriety of an oscillating system. The domain of this statistics starts at the limit of a 
classical coherent state having a Poissonian distributed quanta and may end up with a pure Fock state at the other limit. Studies on the quanta statistics 
had already revealed many pure quantum features for different physical systems and remarkable results were achieved in a large spectrum of photonic 
quantum electrodynamics's (QED) applications \cite{remp} as well as in Bose-Einstein condensate's (BEC) physics \cite{chuu, jacq}. More recently, 
successful experiments in cooling and detection of mechanical systems in the near ground state domain \cite{roch, conn, chan} enhanced a particular 
interest in the research of similar features in the acoustical domain, due to bosonic quantification of the sound's vibrations.

Earlier experiments in laser generation of coherent phonons in different bulk materials \cite{kutt, bart, ezza} were succeeded by new optomechanical 
and electromechanical setups in phonon QED, achieving important experimental results in the acoustical analog of the optical laser by using piezoelectrically 
excited electromechanical resonators \cite{mahb} and laser driven compound microcavities \cite{grud} or trapped ions \cite{vaha}. In the meantime, 
theoretical models propose improvements in the background theory of the experiments like the PT-symmetry approach \cite{jing} and two cavity 
optomechanics \cite{wang}, as well as new possible setups using vibrating membranes \cite{wang, wu}, quantum dots embedded in semiconductor lattices 
\cite{kab1, kab2, okuy} and BECs under the action of magnetic cantilever \cite{bhat}. Moreover, quantum features like sub-Poissonian distributed phonon 
fields have been already predicted in optomechanical setups based on vibrating mirrors \cite{roqu, nati} and in single-electron transistors \cite{merl} as well 
as phonon antibunching \cite{okuy}, squeezing \cite{roqu, kron} and negative Wigner function of phonon states \cite{qian}. In addition to increasing 
performances of the optomechanical devices \cite{aspe}, reports on the art-of-the-state of acoustical cavities \cite{trig,lanz,soyk} have shown good phonon 
trapping in bulk materials with high cavity quality factors, thus leading to the concept of acoustical analog of photon cavity-QEDs (cQED). 

In this article, we study the model of a coherent phonon generator in a setup consisting of a qubit embedded in an acoustical cavity involving strong 
QD-phonon-cavity coupling regime. The qubit, i.e. a two-level quantum dot, is driven by an intense laser field and acts as a phonon source. Under action 
of laser light the electron jumps from the QD's valence band to the conductance band leaving a “hole” in the valence band. The created exciton (electron-hole) 
represents the QD's excited state and interacts with the acoustical vibrations, thus, creating or annihilating phonons in the cavity. We demonstrate that in 
analogy with recent experiments in photon cQED \cite{kim} the strong qubit-resonator couplings may reveal additional quantum phenomena in the phonon 
cQED's domain. Particularly, we show that in this regime the generated steady-state phonon field obeys sub-Poissonian phonon statistics.

This paper is organized as follows. In Sec. II a detailed description of the model is given, i.e., after an introduction of the used nomenclature we focus 
on simplifying the system's Hamiltonian in order to arrive at an easy solvable master equation for the reduced density operator of the QD-phonon system. 
In Sec. III, we present a general overview over the model's results and we further discuss the important aspects of the study. The Summary is given in 
Sec. IV.
\section{Theoretical framework}
\begin{figure}[t]
\centering
\includegraphics[width= 8.6cm ]{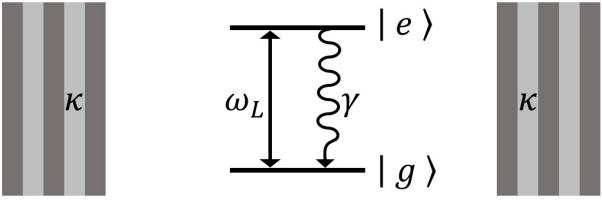}
\caption{\label{fig} 
(color online) The schematic of the investigated model: A two-level QD is fixed in a multilayered structure forming the acoustical nanocavity. 
The QD is pumped near resonance with a coherent laser source of frequency $\omega_{L}$. The emitter may spontaneously emit a photon 
at a decay rate $\gamma$ while the cavity phonon damping rate is denoted by $\kappa$ (see also Ref.~\cite{soyk}).}
\end{figure}

A two-level laser-pumped semiconductor quantum dot is embedded in an acoustical nanocavity (see also \cite{kab1,kab2}). 
The QD's transition frequency  between its ground state $\vert g \rangle $ and 
the excited state $\vert e \rangle$ is denoted by $\omega_{qd}$. The excited QD may spontaneously emit a photon with $\gamma$ 
being the corresponding decay rate, see Fig.~\ref{fig}. For a more realistic case, we introduce the dephasing losses rate through $\gamma_{c}$. 
The single mode cavity phonons of frequency $\omega_{ph}$ are described by the anihilation $(b)$ and the creation $(b^{\dagger})$ operators, 
respectively. The system's Hamiltonian is:
\begin{eqnarray}
H&=& \hbar\omega_{qd}S_{z} + \hbar\omega_{ph} b^{\dagger}b + \hbar \Omega(S^{+}e^{-i \omega_{L} t} + S^{-}e^{i \omega_{L} t}) \nonumber \\
&+& \hbar g S^{+} S^{-} (b^{\dagger} + b),  \label{Htot}
\end{eqnarray}
where the QD's operators are defined as: $S^{+}=\vert e \rangle\langle g \vert $, $S^{-}=\vert g \rangle\langle e \vert $ and 
$S_{z} = \frac{1}{2} \left(  \vert e \rangle\langle e \vert - \vert g \rangle\langle g \vert \right)$ obeying the standard commutation relations for SU(2) 
algebra. The first two terms correspond, respectively, to the unperturbed QD and to the free single mode phonon Hamiltonians. The third term corresponds 
to the QD-laser interaction within rotating wave and dipole approximations whereas $\omega_{L}$ is the laser frequency. The last term describes the 
QD-phonon-cavity interaction with $g$ being the coupling strength constant.

In what follows, we describe the Hamiltonian in a frame rotating at the laser frequency $\omega_{L}$ and apply the dressed-state 
transformation:
\begin{eqnarray}
 \vert + \rangle &=& \sin{ \theta}\vert g \rangle + \cos{\theta}\vert e \rangle, \nonumber \\
 \vert - \rangle &=& \cos{\theta} \vert g \rangle - \sin{\theta} \vert e \rangle ,
 \label{Dstate}
\end{eqnarray}
where $2\theta=\arctan{( 2 \Omega / \Delta )}$ while $\Delta = \omega_{qd} - \omega_{L} $ is the detuning of the laser from the 
QD's transition frequency.  The dressed-state system's Hamiltonian becomes then:
\begin{eqnarray}
H&=&  \hbar \bar{\Omega}R_{z} + \hbar\omega_{ph} b^{\dagger}b + \hbar g(b^{\dagger} + b)\lbrace \sin^2{\theta} \, R_{--} \nonumber \\
&+& \cos^2 {\theta} \, R_{++} - \frac{\sin {(2 \theta)}}{2}  (R^{+} + R^{-}) \rbrace ,
 \label{HDstate}
\end{eqnarray}
where $\bar{\Omega}=\sqrt{\Omega^2 + ({\Delta/2})^2}$. The new QD's operators are: $R^{+}=\vert + \rangle\langle - \vert $, 
$R^{-} = \vert - \rangle\langle + \vert$, $ R_{++} = \vert + \rangle\langle + \vert $, $ R_{--} = \vert - \rangle\langle - \vert $ and 
$R_{z} = R_{++} - R_{--}$ satisfying the commutation relations $ [R^{\pm},R^{\mp}]= \pm R_{z} $ and $ [R^{\mp},R_{z}]=\pm 2 R^{\mp}$.
Again, one performs an unitary transformation to the system's Hamiltonian, $U(t)= e^{i(\bar{\Omega} R_{z}+\omega_{ph} b^{\dagger} b )t}$,
and represents it as follows:
\begin{eqnarray}
H&=& H_{slow} + H_{fast}, \nonumber\\
H_{slow}&=& -\hbar g \frac{\sin{(2 \theta)}}{2} \lbrace b^{\dagger} R^{-} e^{i(\omega_{ph} - 2 \bar{\Omega})t} + \text{H.c.}\rbrace,  \nonumber\\
H_{fast}&=& \hbar g ( \sin^2{\theta}R_{--}+\cos^2{\theta} R_{++} ) \lbrace b^{\dagger} e^{i\omega_{ph}t} + \text{H.c.}\rbrace \nonumber\\
&-& \hbar g \frac{\sin{(2 \theta)}}{2} \lbrace b^{\dagger} R^{+} e^{i(\omega_{ph} + 2 \bar{\Omega})t} + \text{H.c.}\rbrace.  \label{Hspleet}
\end{eqnarray}
Instead of making the usual secular approximation \cite{zb_sc,kmek} we do keep the fast rotating terms in the QD-phonon-cavity 
interaction Hamiltonian. Their main contribution is evaluated  as \cite{jame,gxl}:
\begin{eqnarray}
H_{fast}^{eff} &=& -\frac{i}{\hbar} H_{fast}(t) \int{dt'\, H_{fast}(t')} \nonumber\\
&=& H_{0} -\hbar \bar{\Delta} R_{z} + \hbar \beta b^{\dagger} b R_{z} , \label{Hfast}
 \end{eqnarray}
with
$$\bar{\Delta} = \frac{g^2}{2} \left( \frac{\cos{(2 \theta )}}{\omega_{ph}} - \frac{\sin^2{(2 \theta )}}{4(\omega_{ph} + 2 \bar{\Omega})} \right),$$ 
and 
$$\beta = g^2  \frac{\sin^2{(2 \theta )}}{4(\omega_{ph} + 2 \bar{\Omega})}.$$
Here, $H_{0} $ is a constant and can be dropped as it does not contribute to the system's dynamics. 
Notice that the coupling regimes are related not only to the QD-phonon-cavity coupling constant $g$ but also to the contribution coming from fast rotating terms, 
i.e. $\bar{\Delta}$ and $\beta$, proportional to $g^{2}$. Consequently, for weak QD-phonon couplings strengths $g$ the fast rotating terms contribution 
can be neglected, i.e., $\{\bar{\Delta},\beta\} =0$. For strong QD-phonon coupling regimes the secular approximation is no longer justified and the contribution 
of $H_{fast}^{eff}$ plays a role which will be considered. Thus, the final Hamiltonian $H=H_{slow}+H_{fast}^{eff}$ is: 
\begin{eqnarray}
H &=& \hbar (\omega_{ph} - 2 \bar{\Omega})b^{\dagger} b  -\hbar \bar{\Delta} R_{z} + \hbar \beta b^{\dagger} b R_{z} \nonumber\\
 &-& \hbar g \frac{\sin{(2 \theta)}}{2} \left( b^{\dagger} R^{-} + R^{+} b   \right). \label{Hfinal}
\end{eqnarray}

To solve the QD-phonon system's dynamics one uses the density matrix formalism for the reduced density operator $ \rho $:
\begin{eqnarray}
\dot{\rho} &=& - \frac{i}{\hbar} [H , \rho] + \mathcal{L}_{qd} \rho + \mathcal{L}_{ph} \rho,
 \label{Meq}
\end{eqnarray}
where the Liouville superoperators $ \mathcal{L}_{qd}$ and $ \mathcal{L}_{ph}$ describe, respectively, the QD's and phonons' dissipative effects.
In the bare state representation QD's dissipation processes are expressed by the following spontaneous emission term: 
$\mathcal{L}_{qd} \rho = -\gamma  [ S^{+}, S^{-} \rho ] -\gamma_{c} [ S_{z}, S_{z} \rho ] + \text{H.c.}$ \cite{zb_sc,kmek}. 
In the dressed-state basis and within the secular approximation, i.e. $2\bar \Omega \gg \gamma$, the same processes are described by three 
terms determined by the QD's dressed-state decay rates: 
$\gamma_{+} = \gamma \cos^4 \theta + \frac{1}{4} \gamma_{c} \sin^2{(2 \theta)}$,  
$\gamma_{-} = \gamma \sin^4 \theta + \frac{1}{4} \gamma_{c} \sin^2{(2 \theta)}$ and 
$\gamma_{0} = \frac{1}{4} [ \gamma \sin^2{(2 \theta)} + \gamma_{c} \cos^2{(2 \theta)} ]$. Therefore,
\begin{eqnarray}
\mathcal{L}_{qd} \rho &=& -\gamma_{+}  [ R^{+}, R^{-} \rho ] 
 -\gamma_{-}  [ R^{-}, R^{+} \rho ]  \nonumber\\
& & -\gamma_{0}  [ R_{z}, R_{z} \rho ] + \text{H.c.} .
 \label{LossQD}
\end{eqnarray}
The phonons from the multilayered  acoustical cavity are allowed to interact with the environmental thermal reservoir. In the rotating 
wave approximation this process is described by two terms corresponding respectively to the cavity's damping and pumping effects 
at a rate proportional to $\kappa = \omega_{ph}/Q$ determined by the cavity's quality factor $Q$ \cite{zb_sc}:
\begin{eqnarray}
\mathcal{L}_{ph} \rho &=& -\kappa (1+ \bar{n}) [ b^{\dagger}, b \rho ] -\kappa \bar{n} [ b, b^{\dagger} \rho ] + \text{H.c.}. \label{LossPH}
\end{eqnarray}
Here, $\bar n$ is the mean thermal phonon number corresponding to frequency $\omega_{ph}$ and environmental temperature $T$.

Once the master equation is determined by Eqs.~(\ref{Hfinal})-(\ref{LossPH}), it is solved by projecting the density operator firstly in the 
QD's basis and then in the phonon field's basis \cite{quang}. The projection in the QD's dressed state basis leads after some rearrangements 
to a system of six coupled differential equations involving the following variables: 
\begin{eqnarray}
\rho^{(1)} &=& \rho_{++} + \rho_{--}, \; \; \; \; \,\, \; \rho^{(2)} = \rho_{++} - \rho_{--} , \nonumber\\
\rho^{(3)} &=& b^{\dagger}\rho_{+-} - \rho_{-+} b ,\:  \: \rho^{(4)} = b^{\dagger}\rho_{+-} + \rho_{-+} b , \nonumber\\
\rho^{(5)} &=& \rho_{+-} b^{\dagger} - b \rho_{-+}, \:  \: \rho^{(6)} = \rho_{+-} b^{\dagger} + b \rho_{-+} ,
 \label{MeqQD}
\end{eqnarray}
where $ \rho_{i,j}= \langle i \vert \rho \vert j \rangle,$ $\{ i,j\in \vert + \rangle, \vert - \rangle \}$, are the QD's density matrix elements. 
Finally, the projection in the phonon Fock states basis $ \lbrace \vert n \rangle \rbrace $ gives a set of infinite differential equations, namely,
\begin{eqnarray}
\dot{P}_{n}^{(1)} &=& i g \frac{\sin(2 \theta)}{2} \left( P_{n}^{(3)}-P_{n}^{(5)}  \right) \nonumber\\
&-& 2 \kappa (1+\bar{n})\left(n  P_{n}^{(1)} -(n+1) P_{n+1}^{(1)} \right)  \nonumber\\
&-& 2 \kappa \bar{n}\left( (n+1) P_{n}^{(1)} - n  P_{n-1}^{(1)}  \right),
 \label{MeqQDph1}
\end{eqnarray}
\begin{eqnarray}
\dot{P}_{n}^{(2)} &=& -i g \frac{\sin(2 \theta)}{2} \left( P_{n}^{(3)}+P_{n}^{(5)}  \right) \nonumber\\
&-& 2(\gamma_{+} - \gamma_{-}) P_{n}^{(1)} -  2(\gamma_{+} + \gamma_{-}) P_{n}^{(2)} \nonumber\\
&-& 2 \kappa (1+\bar{n})\left(n  P_{n}^{(2)} -(n+1) P_{n+1}^{(2)} \right)  \nonumber\\
&-& 2 \kappa \bar{n}\left( (n+1) P_{n}^{(2)} - n  P_{n-1}^{(2)}  \right),
 \label{MeqQDph2}
\end{eqnarray}
\begin{eqnarray}
\dot{P}_{n}^{(3)} &=&   i g n \frac{\sin(2 \theta)}{2} \left(  P_{n}^{(1)} -  P_{n}^{(2)} - P_{n-1}^{(1)} -  P_{n-1}^{(2)}\right) \nonumber\\
&-& i \left(  \beta(2 n -1) -\delta \right) P_{n}^{(4)} - (\gamma_{+} + \gamma_{-} + 4 \gamma_{0})  P_{n}^{(3)} \nonumber\\
&-& \kappa (1+\bar{n})\left( (2 n -1) P_{n}^{(3)} - 2(n+1) P_{n+1}^{(3)} + 2 P_{n}^{(5)} \right)  \nonumber\\
&-& \kappa \bar{n}\left( (2n+1) P_{n}^{(3)} - 2 n  P_{n-1}^{(3)}  \right) ,
 \label{MeqQDph3}
\end{eqnarray}
\begin{eqnarray}
\dot{P}_{n}^{(4)} &=& - i\left( \beta (2 n -1) -\delta \right) P_{n}^{(3)} - (\gamma_{+} + \gamma_{-} + 4 \gamma_{0})  P_{n}^{(4)} \nonumber\\
&-& \kappa (1+\bar{n})\left( (2 n -1) P_{n}^{(4)} - 2(n+1) P_{n+1}^{(4)} + 2 P_{n}^{(6)} \right)  \nonumber\\
&-& \kappa \bar{n}\left( (2n+1) P_{n}^{(4)} - 2 n  P_{n-1}^{(4)}  \right),
 \label{MeqQDph4} 
\end{eqnarray}
\begin{eqnarray}
\dot{P}_{n}^{(5)} &=& - i g (n+1) \frac{\sin(2 \theta)}{2} \left(  P_{n}^{(1)} +  P_{n}^{(2)} - P_{n+1}^{(1)} +  P_{n+1}^{(2)}\right) \nonumber\\
&-& i\left( \beta (2 n +1)- \delta \right)P_{n}^{(6)} - (\gamma_{+} + \gamma_{-} + 4 \gamma_{0})  P_{n}^{(5)} \nonumber\\
&-& \kappa (1+\bar{n})\left( (2 n +1) P_{n}^{(5)} - 2(n+1) P_{n+1}^{(5)} \right)  \nonumber\\
&-& \kappa \bar{n}\left( (2n+3) P_{n}^{(5)} - 2 n  P_{n-1}^{(5)} - 2 P_{n}^{(3)} \right) ,
 \label{MeqQDph5}
\end{eqnarray}
\begin{eqnarray}
\dot{P}_{n}^{(6)} &=& - i \left( \beta (2 n +1)-\delta \right)P_{n}^{(5)} - (\gamma_{+} + \gamma_{-} + 4 \gamma_{0})  P_{n}^{(6)} \nonumber\\
&-& \kappa (1+\bar{n})\left( (2 n +1) P_{n}^{(6)} - 2(n+1) P_{n+1}^{(6)} \right)  \nonumber\\
&-& \kappa \bar{n}\left( (2n+3) P_{n}^{(6)} - 2 n  P_{n-1}^{(6)} - 2 P_{n}^{(4)} \right).
 \label{MeqQDph6}
\end{eqnarray}
Here  $ P_{n}^{(i)} = \langle n \vert \rho^{(i)} \vert n \rangle  $ while $\delta = \omega_{ph} - 2\bar{\Omega} + 2\bar{\Delta}$. 

In the next Section we shall describe the cavity phonon dynamics in the steady-state via second-order phonon-phonon correlation function
as well as the mean phonon number.
\begin{figure}[t]
\centering
\includegraphics[width= 8.6cm]{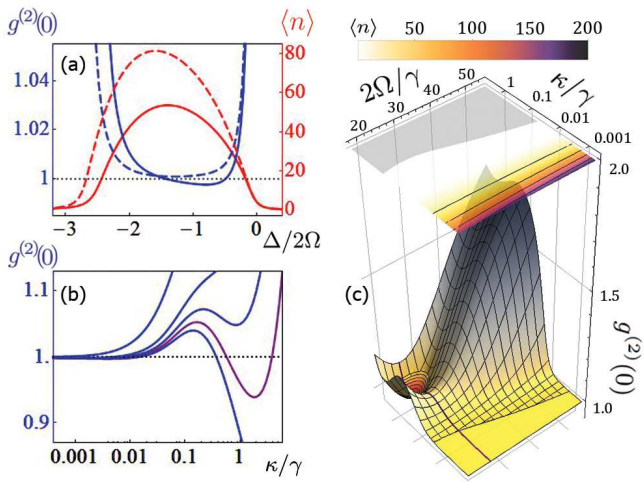}  
\caption{\label{fig1} 
(color online) (a) The second-order phonon-phonon correlation function $g^{(2)}(0)$ (blue lines) and the mean phonon number in the cavity 
$\langle n \rangle$ (red curves) as a function of the QD-laser detuning $\Delta$ normalized by the Rabi frequency $2\Omega$ and beyond 
the secular approximation (continuous lines), and within the secular approximation (dashed curves). Here $ \bar{n} = 0.04 $, 
$2\Omega/\gamma = 25$ and $\kappa/\gamma=5\times 10^{-3}$. (b) $g^{(2)}(0)$ as a function of the normalized damping rate 
$\kappa/\gamma$ for thermal baths at different temperatures and for $2\Omega/\gamma = 25$, and $\Delta/(2\Omega)=-0.7$. Here, 
from top to down one has $\bar{n} = 0.64$, $\bar{n} = 0.16$, $\bar{n} = 0.08$, $\bar{n} = 0.04$ and $\bar{n} = 0.01$, respectively. 
(c) $g^{(2)}(0)$ (3D surface) and $\langle n \rangle$ (density plot) as functions of $\kappa/\gamma$ and $2\Omega/\gamma$. Here 
$\bar{n} = 0.04$ whereas $\Delta/(2\Omega) = -0.7$. The plot regions corresponding to $g^{(2)}(0) >1$ and $g^{(2)}(0)<1$ are 
represented in different mesh styles.  The plot region corresponding to $\langle n \rangle < 1$ is represented in gray color. The purple 
curves in figures (b) and (c) are identical. Other parameters are: $\gamma_{c}/\gamma = 0.1$, $g/\gamma=15$, $\omega_{ph}/\gamma = 35$.}
\end{figure}
\begin{figure}[b]
\centering
\includegraphics[width= 8.6cm ]{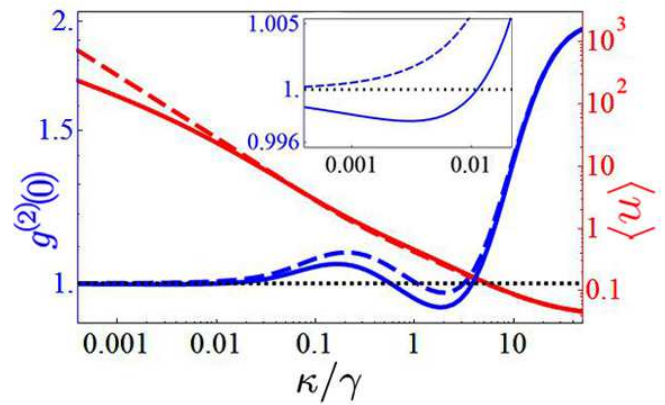}
\caption{\label{fig2} 
(color online) The second-order phonon-phonon correlation function $g^{(2)}(0)$ (blue lines) and the mean phonon number in the cavity 
mode $\langle n\rangle$ (red curves) as a function of the cavity damping rate $\kappa$ normalized by the spontaneous emission rate 
$\gamma$. The continuous curves are beyond the secular approximation whereas the dashed ones are within the secular approximation. 
Here, $2\Omega/\gamma = 25$, $\Delta/(2\Omega)=-0.7$, $\bar{n}=0.04 $, $\gamma_{c}/\gamma=0.1$, $g/\gamma=15 $ and 
$\omega_{ph} /\gamma = 35$. The continuous blue line is identical with the purple lines of Fig.~(\ref{fig1}). The inset picture represents 
a close look on the behaviors of the second-order correlation functions in the regions around $10^{-3} \leq \kappa/\gamma \leq 10^{-2}$. }
\end{figure}

\section{Results and Discussion}
Consequently, the average phonon number in the cavity mode is expressed as:
\begin{eqnarray}
\langle n \rangle &=& \langle b^{\dagger} b \rangle = \sum_{n=0}^{\infty} n P_{n}^{(1)}. \label{n}
\end{eqnarray}
The nanocavity second-order phonon-phonon correlation function is defined as usual \cite{glau},
\begin{eqnarray}
g^{(2)}(0) &=& \frac{ \langle b^{\dagger} b^{\dagger} b b \rangle }{ \langle b^{\dagger} b \rangle ^{2} } 
= \frac{\sum_{n=0}^{\infty} n (n-1) P_{n}^{(1)}}{\langle n \rangle ^{2}}. \label{g2}
\end{eqnarray}
The system of Eqs.~(\ref{MeqQDph1}) - (\ref{MeqQDph6}) as well as the infinite series in the expressions (\ref{n}) - (\ref{g2}) must be 
truncated at a particular value of $n=N_{max}$ such that the variables of interest remain unchanged if one further increases $N_{max}$ \cite{ginz}.

In what follows, we shall study the system in the steady-state regime, i.e., $ \dot{P}_{n}^{(i)}=0 $ for $ \{i = 1 \cdots 6 \}$. 
The second-order correlation function given by  Eq.~(\ref{g2}) and the average phonon number in  Eq.~(\ref{n}) are used to describe the 
phonon field behaviors in the acoustical cavity mode \cite{glau}. Once truncated, the system of coupled equations 
(\ref{MeqQDph1}) - (\ref{MeqQDph6}) is solved by setting the model's parameters 
$\{\gamma , \gamma_{c}, g, \omega_{ph}, \kappa, \bar{n}, \Omega \}$ and $\Delta$, respectively.

A general overview over the steady-state system's behavior in the strong coupling regime or within the secular approximation, i.e. when
$\{\bar \Delta, \bar \beta = 0\}$, is presented in Fig.~\ref{fig1}. The phonon field's statistics is described by the second-order correlation 
function $g^{(2)}(0)$ with $g^{(2)}(0) = 1$ describing the Poissonian phonon distribution while when $g^{(2)}(0) < 1$ one has sub-Poissonian 
phonon statistics. One observes that moderate laser-QD coupling strengths, i.e. $\Omega$, as well as lower temperatures give a more 
prominent sub-Poissonian phonon statistics. Furthermore, beyond the secular approximation, the phonon statistics exhibits quantum 
features, i.e. $g^{(2)}(0)<1$ (compare the corresponding curves in Fig.~\ref{fig1}a) and the effect is more pronounced for stronger 
QD-phonon coupling strengths. Fig.~\ref{fig1}(b) shows, beyond the secular approximation, the second-order phonon-phonon correlation 
function as a function of $\kappa/\gamma$. Here, again, around $\kappa > \gamma$ one has a quantum phonon effect, i.e. sub-Poissonian 
phonon statistics. Thus, the contribution to the system's dynamics of the fast rotating terms $H_{fast}^{eff}$ evaluated by the Eq.~(\ref{Hfast}) 
are essential for a sub-Poissonian quantum feature, see Fig.~\ref{fig1}. In Fig.~\ref{fig2} we compare our result within and beyond the 
secular approximation. The second-order correlation function estimated in both cases converges for higher and smaller cavity damping rates. 
However, for lower values of $\kappa/\gamma$ quantum features are proper only beyond the secular approximation 
(see the inset picture in Fig.~\ref{fig2}). Furthermore, the mean phonon number decreases in this particular case although it is quite high comparing 
to the $\kappa > \gamma$ situation.

\section{Summary}
In summary, we have investigated the phonon quantum statistics in an acoustical nanocavity. A laser-pumped two-level quantum dot 
is embedded in the cavity contributing to phonon quantum dynamics. We have demonstrated that stronger QD-phonon-cavity coupling regimes 
lead to quantum features of the cavity phonon field in the steady-state. This QD-cavity interaction regime obliges to go beyond the 
secular approximation. Ignoring this fact would lead to an erroneous estimation of the phonon statistics for some parameter domains. 
\section*{Acknowledgement}
We are grateful to financial support via the research Grant No. 13.820.05.07/GF.


\end{document}